\documentstyle[aps,graphics,epsf,twocolumn]{revtex}

\begin{document}
\draft
\title{Diffusion of single long polymers in fixed and low density matrix of obstacles confined to two dimensions}
\author{Ryuzo Azuma and Hajime Takayama}
\address{Institute for Solid State Physics, The university of Tokyo\\
7-22-1 Roppongi, Minato-ku, Tokyo 106-8666, Japan}
\date{\today}
\maketitle
%\begin{multicols}{2}
%\input{intro.tex}
%\input{method.tex}
%\input{theory.tex}
%\input{result.tex}
%\input{dissc.tex}
%\input{summary}
%\input{acknow.tex}
%\appendix
%\input{appendi2.tex}
\begin{abstract} 
%\parbox[c]{\linewidth}{
Diffusion properties of a self-avoiding polymer embedded in 
regularly distributed obstacles with spacing $a=20$ and confined in 
two dimensions is studied numerically using the extended bond 
fluctuation method which we have developed recently.
We have observed for the first time to our knowledge, that the mean square
 displacement of a center monomer 
$\phi_{M/2}(t)$ exhibits four dynamical regimes, i.e., 
$\phi_{M/2}(t) \sim t^{\nu_m}$ with $\nu_m\sim 0.6$, $3/8$, $3/4$, 
and $1$ from the shortest to longest time regimes. The exponents in 
the second and third regimes are well described by segmental diffusion 
in the ``self-avoiding tube''. In the fourth (free diffusion) regime, we have 
numerically confirmed the relation between the reptation time $\tau_d$
and the number of segments $M$, $\tau_d\propto M^3$. 
%}
\end{abstract}
%\def\nle{\ \raise.3ex\hbox{$<$}\kern-0.8em\lower.7ex\hbox{$\sim$}\ }
%\def\nge{\ \raise.3ex\hbox{$>$}\kern-0.8em\lower.7ex\hbox{$\sim$}\ }

%\documentstyle[11pt,fullpage]{article}
%\setlength{\textheight}{240mm}
%\setlength{\textwidth}{160mm}
%\setlength{\topmargin}{-24mm}

%\begin{document}
\section{Introduction}

The reptation theory, proposed for concentrated solution of linear 
polymers~\cite{deGennes-71}, is based on the assumptions that a 
polymer moves in a ``tube'' made of the other polymers and that the 
tube does not deform till the polymer of interest creeps out of it. 
Since the theory was proposed, various investigations have been done 
on dynamic properties of a polymer in networks of immobile obstacles. 
As for the case of concentrated solution, Baumg\"artner et 
al~\cite{Baumgartner-1} studied dynamics of a freely jointed chain 
with the Lennard-Jones potential in a dense network of frozen-in 
chains. Evans and Edwards~\cite{evans}, on the other hand, proposed 
an efficient lattice model with stochastic local jump motion of 
segments in regularly distributed obstacles on the cubic lattice, 
which simulates such phenomena as dynamics of a DNA molecule in 
agarose gel. 

Study of diffusion property of a single real polymer (self-avoiding 
chain) in fixed and {\it low density} network of obstacles is of importance,
 since details of DNA dynamics in agarose gel, 
etc.~\cite{Norden,ZimmLev,Austin} have been studied by
many experimental works along with recent developments in technical aspects
\cite{Yanagida-82,Bustamante-91}.
To see entanglement effect, investigations of a self-avoiding polymer 
in two dimensions would be desired. In two dimensions, a molecule with relatively 
low polymerizations can entangle with many obstacles since it extends 
larger than a polymer in three dimensions.

Recently, Maier and R\"{o}dler~\cite{Maier} observed conformation and 
diffusion of a single DNA molecule electrostatically bound to fluid 
cationic lipid bilayers by fluorescence microscopy. They measured 
radius of gyration $R_I$ and obtained the relation 
$\langle R_I^2\rangle \sim N^{2\nu}$ where $N$ is the number of 
base pairs of the molecule and $\nu = 0.79\pm0.04$. For diffusion 
coefficient of the center of mass $D_G$ they showed $D_G\propto N^{-1}$. 
These results agree well with the theoretical predictions for a 
self-avoiding polymer in two dimensions without hydrodynamic 
interactions. 
It seems feasible to arrange obstacles on that layer and somehow 
measure displacement of a particular part of DNA to examine diffusion 
property of a self-avoiding polymer in a fixed network of obstacles 
in two dimensions. 

In the present work, dynamic properties of a single polymer in 
regularly distributed obstacles in two dimensions is studied by means 
of the extended bond fluctuation method (e-BFM) which we have developed 
recently~\cite{AT-1}. From the simulated mean square displacement of a 
center monomer $\phi_{M/2}(t)$, we can distinguish clearly four 
dynamical regimes, as is the case of a Gaussian chain~\cite{DoiEdwa} 
in the same network of obstacles. The simulated exponents $\nu_m$ of 
$\phi_{M/2}(t) \sim t^{\nu_m}$ are $\nu_m\sim 0.6$, $3/8$, $3/4$, and 
$1$ from the shortest to longest time regimes. The exponents in the 
early three regimes differ from those of the Gaussian chain, but they 
are well described when the excluded volume effect is taken into 
account. In particular, dynamics in the second and third regimes is 
interpreted as segmental diffusion in the ``self-avoiding tube,''
 where displacement $\Delta s$ along 
that tube corresponds to $a\Delta s^\nu$ in real space with $a$ and 
$\nu$ being the tube diameter and the exponent relating radius of 
inertia $R_I$ and the number of segments $M$ as $R_I \sim M^\nu$, 
respectively. The reptation time $\tau_d$, as for the crossover time
 between the third and fourth (free diffusion) regimes, is ascertained to be 
proportional to $M^3$ with $M$ being the number of segments of the 
chain.

In the next section we explain the details of the simulational methodology
and parameters which we have used. 
The scaling theory of tube model for a self-avoiding polymer in the network of
 fixed and low density obstacles is described in the Sec.~\ref{sec:theory}. 
We then present the results and their scaling analyses compared to the theory in
Sec.~\ref{sec:result}, and conclude them in the final section. 
%\cite{kremer-3}
%\cite{paul-2}
%\cite{yamakov}
%\cite{schulz}
%\cite{pincus}
%\cite{bishop}
%\cite{paul}
%\cite{deGennes-2}
%\cite{dbios}
%\cite{deGennes-3}
%\cite{Kremer-1}
\section{Method}

The bond fluctuation method (BFM)~\cite{carkre,carkre-2} is known 
to provide effective Monte Carlo (MC) algorithm to study static 
and dynamic properties of various phenomena involved in 
polymers~\cite{kremer}. We have adopted a square lattice version 
of the method where a polymer is described by a sequence of 
monomers which occupies four sites of the smallest square cell. 
The monomers are not allowed to occupy the same site (excluded 
volume interaction or effect (EVE)) and connected by bonds which 
can vary its length $l$ in the range $2\le l \le \sqrt{13}$. 
The elemental MC step is a trial of displacement of each monomer 
by one lattice spacing to a randomly chosen direction. The process 
satisfies the self-avoiding condition. 

The extended-BFM which we have proposed~\cite{AT-1} and used in 
the present study incorporates a non-local  movement of the 
`s-monomer' into the BFM. 
The s-monomer is defined as the one whose nearest neighbors 
separate not greater than $\sqrt{13}$ with each other. 
An s-monomer can be displaced to any monomer pairs between the 
nearest s-monomers by a stochastic process which fulfills detailed 
balance condition. This new non-local process can be introduced 
in any relative frequency into the conventional-BFM above 
mentioned.  We have used the following combined process: at even 
(odd) time step of the c-BFM, the new process for even (odd) 
numbered s-monomers is tried. 

The e-BFM has been initially introduced to overcome the trapping or pinning 
difficulty which the c-BFM is faced in gel electrophoresis under 
a large electric field. It has turned out that the method 
reproduces static and dynamic properties also under a vanishing 
field correctly, and more faster than the c-BFM for polymer 
dynamics in a space with fixed obstacles. In fact, the ratio $A$ 
of the diffusion constants obtained by the c-BFM and e-BFM is 
less than unity: $A\sim 1/2$ and $1/3$ for $a=\infty$ (no 
obstacles) and $20$, respectively, with $a$ being the distance 
between the obstacles. 

By means of the e-BFM we here mainly observed the mean square displacement
of the center monomer $\phi_{M/2}(t)$ as a representative of inner 
monomers. The initial configurations are provided by the 
self-avoiding walk algorithm where monomer sequence is made by iteration of
 trials of making a new end. If it finds bond crossing or monomer overlap,
it stops making the chain and go to another starting position. 
Although this algorithm can give 
independent equilibrium configurations, the averaged time needed for
 a complete configuration turned out to be quite large due to the low acceptance
rate for each trial. So we have relaxed the condition for accepting a new end as 
that it allows up to $\omega$ trials. Here we have chosen $\omega=7\sim 14$. 
%So we 
%modified it so as to ???????????---->
%make initial shapes which are not necessarily in equilibrium and waited until they relaxed
%in $R_I$ during idle runs.  
The number of samples given at $t=0$, the
 start time of the idle run, was $512\sim 788$. 
The interval of the measuring run $t_f-t_i$ was taken so as to $\left[\phi_{M/2}
(t_f-t_i)\right]^{1/2}\gtrsim CR_I$, with $C \gtrsim 2$. %<------??????????
The system size was $3M\times 3M$ in which even a fully extended 
polymer cannot interact with itself through the periodic boundary 
conditions introduced. We distributed, regularly in a spacing $a=20$, obstacles
 of a unit square cell, with which any monomer site cannot overlap (the EVE). 

\begin{figure}
\leavevmode\epsfxsize=80mm
\epsfbox{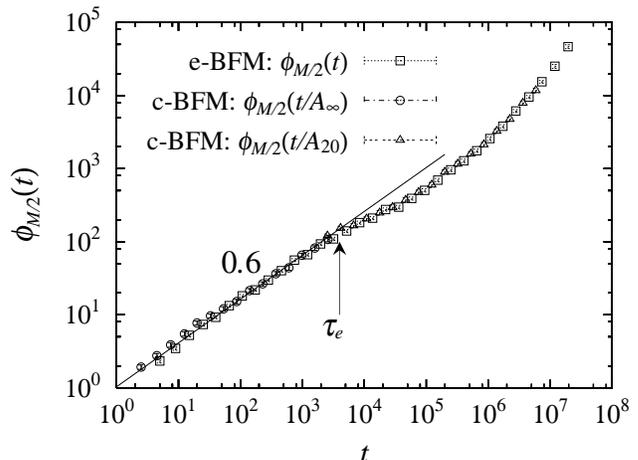}
%\epsfbox{../fig-eps/c-e-BFM.ps}
\caption{
The mean square displacements of the center monomer of the $M=150$ chain 
in regular distributed (spacing $a=20$) obstacles. The symbols $\Box$, $\bigcirc$, and
$\triangle$ are of the e-BFM, c-BFM for $t<\tau_e$, and c-BFM for $t>\tau_e$ respectively. 
}
\label{fig:c-e-BFM}
\end{figure}

%???????????---->
Let us here compare some results simulated by the c-BFM and the e-BFM. 
As an example, we show $\phi_{M/2}(t)$ for $a=20$ and $M=150$ in
 Fig.~\ref{fig:c-e-BFM}. 
At $t<\tau_e$, $\tau_e$ being the crossover time between regimes I and II
 introduced in the next section, $\phi(t/A_\infty)$ with $A_\infty\cong 0.5$
 simulated by the c-BFM and $\phi_{M/2}(t)$ by the e-BFM coincide with each
 other, and exhibit the power-law growth which is proportional to $t^{0.6}$. 
The latter is the characteristics of monomer diffusion in free space. 
At $t>\tau_e$, on the other hand, where monomer diffusion is affected by the 
surrounding obstacles, $\phi_{M/2}(t)$ by the e-BFM is seen to coincide with
 $\phi_{M/2}(t/A_{20})$ with $A_{20}\cong 0.3$ by the c-BFM. The factors
$A_\infty$ and $A_{20}$ are just the ratios $A$ of the diffusion constants mentioned above. 
Thus the results shown in Fig.~\ref{fig:c-e-BFM} demonstrates that, there
is no fundamental difference in diffusion properties of the polymer simulated
 by the c-BFM and the e-BFM.
%It can been seen that for $t<\tau_e$, the monomer diffuses as if it is in free space,
% showing that the relation $\phi_{M/2}(t)\propto t^{0.6}$ holds \cite{carkre}, 
%however, for  $t>\tau_e$, the monomer starts to feel the surrounding obstacles.
%It can be shown that there is no fundamental change in the diffusion property of monomer
% of the e-BFM to the c-BFM.
%If we take into account the above mentioned acceleration constant,
% it can be expected that 
%$\phi_{M/2}(t/A_\infty)$ of the c-BFM for $t<\tau_e$ be just on $\phi_{M/2}(t)$ of the e-BFM
%because the motion of monomer in such time region does not affected by the obstacles
%and thus $A=A_\infty$. 
%But on the other hand, $\phi_{M/2}(t/A_{20})$ of the c-BFM for 
% $t>\tau_e$ is to be just on the curve $\phi_{M/2}(t)$ of the e-BFM
%for the diffusion is faster than the c-BFM by $A=A_{20}$. 
%In fact, such relation between $\phi_{M/2}(t)$ of c-BFM and e-BFM, are confirmed by
% Fig.~\ref{fig:c-e-BFM}. %<------??????????
%\begin{wrapfigure}[20]{i}[0.pt]{8.cm}
%    \scalebox{0.44}{\includegraphics{T20O150.CON-GTM.ps}}
%    \caption{
%The mean square displacements of the center monomer of the $M=150$ chain 
%in regular distributed (spacing $a=20$) obstacles. The symbols $\Box$, $\bigcirc$, and
%$\triangle$ are of the e-BFM, c-BFM for $t<\tau_e$, and c-BFM for $t>\tau_e$ respectively. 
%}
%    \label{fig:T20O150.CON-GTM}
%\end{wrapfigure}
\section{Scaling Theory }\label{sec:theory}

We begin with brief summary of the scaling argument on the tube theory of a 
Gaussian chain moving in a network of fixed obstacles~\cite{DoiEdwa}. 
There exist four time regimes in which the chain exhibits different 
dynamics. In the earliest regime denoted by regime I, the $n$-th inner 
monomer fluctuates without perceiving wall of the tube, and its mean square 
displacement $\phi_n(t)$ is written as $\phi_n(t)\sim t^{\nu_m}$ with 
$\nu_m =1/2$~\cite{deGennes-67}. Here we put $\zeta/k_BT=1$ with 
$\zeta$ being the friction coefficient of a segment and the segment 
length $b$ unity. 
In regimes II and III, the polymer motion perpendicular to 
the tube coordinate is restricted. Diffusional displacement of the 
primitive chain along the tube  $\langle \Delta s^2\rangle$ in a time 
interval $t$ is written as $\langle \Delta s^2\rangle 
\sim at^{1/2}$ and $at^1$ before (regime II) and after (regime III) 
the Rouse relaxation time $\tau_r\sim M^2$, where $a$ is unit length of
the primitive chain which is set equal to
the tube diameter. Such displacement along the tube corresponds to that 
in real space $\langle \Delta x^2\rangle$ through 
\begin{equation}
\langle \Delta x^2\rangle\cong \langle \Delta  s^2\rangle^\nu
\label{eq:dels-delx}
\end{equation}
with $\nu =1/2$ since the tube is Gaussian. 
Thus we obtain $\nu_m = 1/4, 1/2$ for regimes II and III, 
respectively. In regime IV, the whole chain creeps off 
the tube over the length scale of $R_I$, and motion of the segment 
coincides with that of the center of mass which follows the linear relation 
$\phi_n(t)\sim R_I^2t/\tau_d$. Here $\tau_d\sim M^3/a^2$ is the reptation 
time. 

In regime IV of a model simulating a Gaussian chain in the regularly 
distributed obstacles~\cite{evans}, viscosity $\eta$, which is compared 
to $\tau_d$~\cite{DoiEdwa}, was measured up to $L=50$ with $L$ being the 
number of beads and was shown to be proportional to $L^{3.41\pm0.14}$ 
using $L\ge 20$~\cite{deuma-2}. 
Recent study on this model of the most concentrated case clearly showed 
$\nu_m=1/4$ in the shorter time region, i.e. regime II (regime I was not seen)~\cite{ebert}. 
%??????---> Are these 
%studies on a Gaussian chain? <----???????

Now let us turn to dynamics of a real chain (with the EVE) in space with 
fixed obstacles. In regime I, as for a Gaussian chain, inner monomers 
fluctuate without the tube constraint but with the EVE. In this case, or 
without obstacles ($a=\infty$), it is known that $\phi_{M/2}(t) \propto 
t^z$ at $t \ll \tau_r$, where $z=1/(1+(1/2\nu_F))$ with $\nu_F = 3/(d+2)$ 
and $d$ being dimension of the space, and $\tau_r$ is the rotational relaxation
 time for a self-avoiding chain~\cite{Kremer-1}.

\begin{figure}
\leavevmode\epsfxsize=80mm
\epsfbox{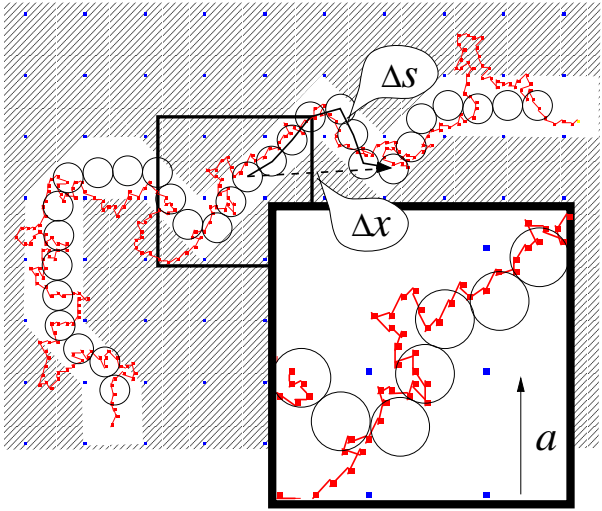}
\caption{
The schematic representation of a chain of blobs. The polymer configuration 
drawn is one of instantaneous ones observed in our simulation with $M=400$.
The blobs are drawn intuitively from the polymer configurations averaged
 over time interval of about $\tau_e$. In an instantaneous configuration there
 some local configuration of hernia type are seen. 
}
\label{fig:bl-chain}
\end{figure}
\begin{figure}
\leavevmode\epsfxsize=80mm
\epsfbox{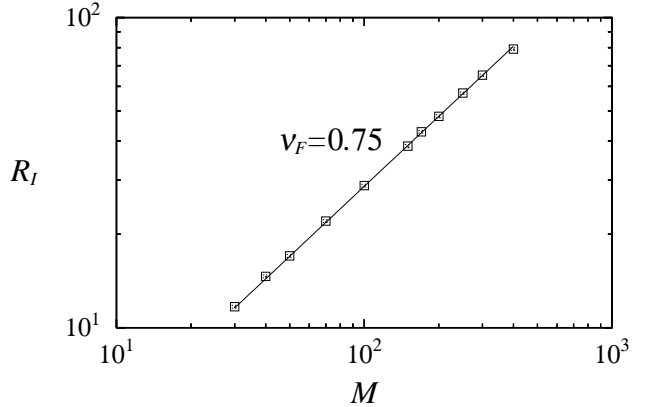}
%\epsfbox{../fig-eps/mnuF.ps}
\caption{
The relation between radius of gyration $R_I$ and length $M$ at $a=20$. 
}
\label{fig:mnuF}
\end{figure}

A polymer configuration in regimes II and III is considered as a chain of 
blobs along the tube coordinate as shown schematically in Fig.~\ref{fig:bl-chain}. 
Each blob of an average size $a_{\rm b}$ consists 
of $g$ ($=a_{\rm b}^{1/\nu_F}$) monomers on average. In terms of the tube 
coordinate, the end-to-end distance $R$ is written $R/a_{\rm b} \sim 
(M/g)^{\nu_F}$, which implies $R_I \propto M^{\nu_F}$ for the original 
chain. In fact the latter has been confirmed numerically at least for low 
concentration of obstacles in two dimensions as shown in Fig.~\ref{fig:mnuF}. 
Thus the EVE plays an essential role in determining
a ``self-avoiding tube'' in these regimes.

For dynamical aspects under a given polymer configuration (and so a 
self-avoiding tube), however, the EVE between the blobs of length scale $a_{\rm b}$ 
is considered to be screened out~\cite{pincus,brochard,deGennes}. Accordingly, 
the diffusion process of blobs along the tube is the same as that of the 
Gaussian chain~\cite{deGennes-71}. Thus, the exponents of the mean square 
displacement of the center monomer along the tube $\phi_{M/2\parallel}(t)$ are given 
by $1/2$ and $1$ for $t<\tau_\theta$ and $t>\tau_\theta$, respectively, 
where $\tau_\theta=\tau_e(M/g)^2$ is the Rouse relaxation time of the chain 
of blobs. Now taking into account eq.~(\ref{eq:dels-delx}) with $\nu = 
\nu_F$, we obtain the exponents $\nu_m=\nu_F/2$ and $\nu_F$ in regimes II 
and III, respectively.  In regime IV $\nu_m=1$ as in the Gaussian chain. 

Now we derive $\phi_n(t)$ with $n\sim M/2$ in each regime more quantitatively. 
As the Rouse and the reptation theory claim relations $\tau_\theta/\tau_e\sim
 \left[M/g\right]^2$ and $\tau_d/\tau_e\sim \left[M/g\right]^3$, 
the pre-factors of $\phi_n(t)$ in the all regimes can be determined by 
matching its scaling forms in the neighboring regimes of $\tau_e$,
 $\tau_\theta$ and $\tau_d$. 
Setting $b,\zeta/k_BT=1$ as before and noting that 
$\phi_n\sim R^2\left[t/\tau_d\right]$ in regime IV, we obtain 
$\phi_n(t)$ and the cross-over times as listed on Table \ref{table:tube-tab}.
%\suppressfloat[b]
%\begin{center}
\enlargethispage{4cm}

\begin{table}%[!tb]
\begin{tabular}[!tb]{|c|c|c|}\hline
 & real chain& Gaussian chain\\\hline
I & $\phi_n\sim t^z$ & $\phi_n\sim t^{1/2}$\\\hline
$\tau_e$ & $\sim a_{\rm b}^{2/z}$ & $\sim a^4$\\\hline
II & $\phi_n\sim a_{\rm b}^2\left[t/\tau_e\right]^{\nu/2}$ & $\phi_n\sim a^2\left[t/\tau_e\right]^{1/4}$\\\hline
$\tau_\theta$ & $\sim \tau_e\left[M/g\right]^2$ & $\sim M^2$\\\hline
III & $\phi_n\sim a_{\rm b}R\left[t/\tau_\theta\right]^\nu$ & $\phi_n\sim aR\left[t/\tau_\theta\right]^{1/2}$\\\hline
$\tau_d$ & $\sim \tau_e\left[M/g\right]^3$ & $\sim M^3/a^2$\\\hline
IV & $\phi_n\sim R^2\left[t/\tau_d\right]$ & $\phi_n\sim R^2\left[t/\tau_d\right]$\\\hline
\end{tabular}
\caption{Expressions of $\phi_n(t)$ in the four regimes
\label{table:tube-tab}}
\end{table}
%\end{center}

Lastly we consider the relaxation time $\tau_D$ of the auto-correlation function of the
end-to-end vector defined by $P(t)=\langle{\bf R}(t){\bf R}(0)\rangle \propto
 \exp\left[-t/\tau_D\right]$. For regime IV the reptation theory \cite{DoiEdwa}
predicts that $\tau_D$, called the reptation or disengagement time, is obtained
from one-dimensional diffusion equation of the primitive chain.
It is given by $\tau_D=L^2/\pi^2D_{\rm c}$, where $D_{\rm c}=1/M$ is the diffusion
 constant along the tube, and $L$ is the length of the primitive chain and is given
 by $aL=M$ for a Gaussian chain ( note we put $k_BT/\zeta=1$ and $b=1$ ).
We can rewrite $\tau_D$ for a self avoiding chain by replacing $L$ by $a_{\rm b}M/g$.
This is because the motion of the chain along the tube is described by dynamics of a
sequence of blobs which behave as a Rouse chain. Therefore $\tau_D$ and $\tau_d$ in
 Table~\ref{table:tube-tab} are related as
\begin{equation}
\tau_D=\frac{1}{\pi^2}\left[\frac{R}{R_I}\right]^{3/\nu_F}\tau_d \cong3.65\tau_d\label{eqn:tau_dD}
\end{equation}
for a real chain. 
\section{Results }\label{sec:result}

\begin{figure}
\leavevmode\epsfxsize=80mm
\epsfbox{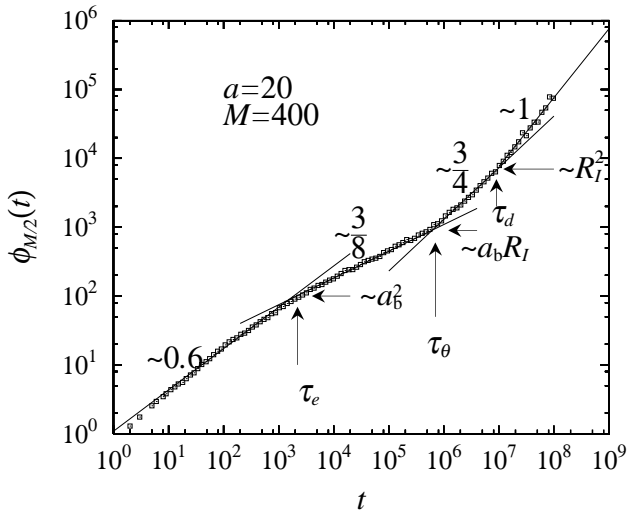}
%\epsfbox{L400-x5.ps}
\caption{
The mean square displacement of the center monomer of the $M=400$ chain 
in regular distributed (spacing $a=20$) obstacles. 
}
\label{fig:L400-x5}
\end{figure}

In Fig.~\ref{fig:L400-x5} we show the result of the mean square displacement 
of the center monomer $\phi_{M/2}(t)$ obtained from our longest chain $M=400$ 
in two dimensions. As listed on Table 1 in the previous section, 
$\phi_{M/2}(t)$ is expected to exhibit different power-law dependences on 
$t$ in the four regimes: with $d=2$ and so $\nu_F=3/4$ and $z=3/5$, $\nu_m = 
3/5, 3/8, 3/4$ and $1$ in from regime I to IV, respectively. 
These theoretical results are well reproduced in our simulated results in the 
figure. The auxiliary line in the shortest time interval (regime I) stands 
for the least square fit of our separately simulated data of the $M=400$ 
chain with $a=\infty$ (no obstacles) to the power-law function of $t$. 
The exponent is estimated as $\nu_m=0.60\pm 0.01$. 
The data of $a=20$ lie upon this line at $10 \lesssim t \lesssim 2\times10^3$. 
 as well. 
%\begin{wrapfigure}[20]{i}[0.pt]{8.cm}
%    \scalebox{0.44}{\includegraphics{T20O400-x5.ps}}
%    \caption{
%The mean square displacement of the center monomer of the $M=400$ chain 
%in regular distributed (spacing $a=20$) obstacles. 
%}
%    \label{fig:T20O400-x5}
%\end{wrapfigure}

The tube model predicts the crossover time between regimes I and II, 
where fluctuating monomers start to feel the wall of tube and to form blobs, as 
$\left(t, \phi_{M/2}(t)\right)=\left(\tau_e,a_{\rm b}^2\right)$. 
From inspection of the figure, its ordinate seems significantly smaller as compared to 
$a^2$. It is estimated as $a_{\rm b}\sim a/2$. 

We may think of another diameter $d_T$, which corresponds to diameter of a straight
 tube with hard wall. The mean square displacement of a
 monomer perpendicular to tube $\phi_{M/2\bot}(t)$ for $t\gg \tau_e$ is estimated as
\begin{eqnarray}
\phi_{M/2\bot}(t\gg \tau_e)
\cong& \frac{\displaystyle\sum^{d_T-2}_{i=1}\sum^{d_T-2}_{j=1}\left[i-j\right]^2}{\displaystyle\sum^{d_T-2}_{i=1}\sum^{d_T-2}_{j=1}}\nonumber\\
=&\frac{1}{6}\left[d_T-1\right]\left[d_T-3\right].\label{eqn:d_T}
\end{eqnarray}
Replacing $\phi_{M/2\bot}(t)$ by $\phi_{M/2}(\tau_e)=a_{\rm b}^2=114\pm 5$ and
substituting it to eq.~(\ref{eqn:d_T}), we get $d_T\simeq 28.2\pm0.57$. 
This indicates that the tube diameter $d_T$ is bigger than $a$ since the ``tube''
 is not a tube with hard wall but is formed in a space of the point obstacles. 
%This is qualitatively in agreement with the result for a long chain in a 
%straight tube~\cite{Kremer-1}, where the perpendicular component of the 
%square displacement saturates to about $a/4$ at $t>\tau_e$. 
%??????????-----> This indicates that the monomer feels the potential of 
%the obstacles with infinite
% value and tend to place in a space where distance from the wall is
% $\lambda\sim \left[a-a_{\rm eff}\right]/2$. 
%Thus it is reasonable that $\lambda/a$ of the gel is lesser than that of 
%the straight tube. <-----??????????????

The data in Fig.~\ref{fig:L400-x5} also demonstrate nicely the chain dynamics 
in the ``self-avoiding tube" described in the previous section. In fact 
the predicted exponents $3/8$ and $3/4$ in regimes II and III respectively 
are ascertained with the appropriate crossover point 
$\left(t, \phi_{M/2}(t)\right)=\left(\tau_\theta,a_{\rm b}R_I\right)$. 
Finally at $t \gtrsim \tau_d$ crossover from region III to IV occurs, and 
the linear relation $\phi_{M/2}(t) \sim t$ is reproduced in regime IV.

\begin{figure}
\leavevmode\epsfxsize=80mm
\epsfbox{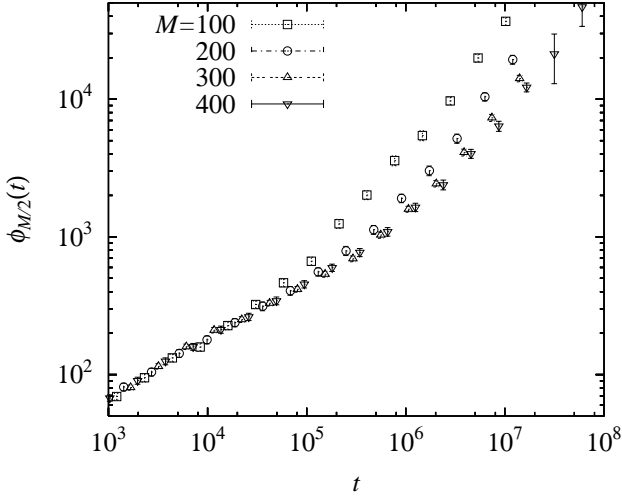}
%\epsfbox{L400-x5.ps}
\caption{
The simulated data of $\phi_{M/2}(t)$ with $M=100$, $200$, $300$ and $400$ at $a=20$.
As expected from the fact $R_I > a_{\rm b}$ for these $M$ they coincide 
with each others in regime I and a shorter time range of regime II. 
}
\label{fig:rawphis}
\end{figure}

In order to confirm further the compatibility of theoretical arguments 
on the polymer dynamics in the self-avoiding tube, we have carried out 
scaling analyses on $\phi_{M/2}(t)$ with different $M\ (=100 \sim 400)$. 
Some raw data are shown in Fig.\ref{fig:rawphis}. 

\begin{figure}
\leavevmode\epsfxsize=80mm
\epsfbox{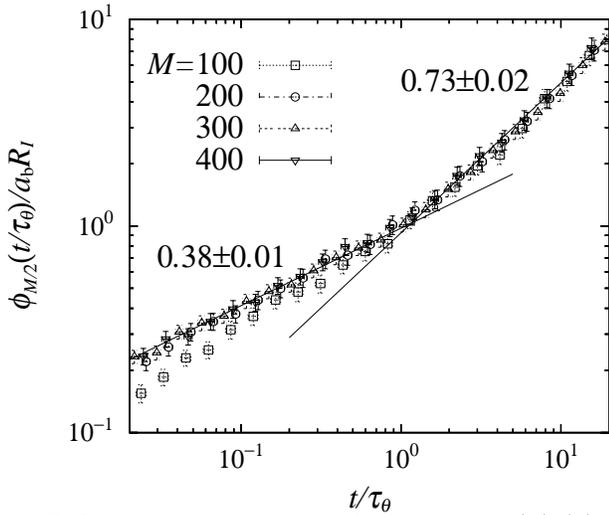}
%\epsfbox{scl2-3.ps}
\caption{
The double logarithmic plot of $\phi_{M/2}(t/\tau_\theta)/a_{\rm b}R_I$ 
with $t/\tau_\theta$. 
}
\label{fig:scl2-3}
\end{figure}

In Fig.~\ref{fig:scl2-3}, double logarithmic plots of 
$\phi_{M/2}(t/\tau_\theta)/a_{\rm b}R_I$ in terms of $t/\tau_\theta$
around crossover between regimes II and III are shown for $M=100$, $200$, $300$ and $400$.
Here the value $\tau_\theta$ is determined as $\tau_\theta = 
\tau_e\left[R_I/a_{\rm b}\right]^{2/\nu_F}$, 
using $R_I$ and $\nu_F$ in Fig.~\ref{fig:mnuF} and $\tau_e$ and $a_{\rm b}$ 
from Fig.~\ref{fig:rawphis}. 
 It can been seen that the scaling fit is satisfactory for $M\ge 200$. The 
auxiliary lines fitted to the data yield exponents $0.38\pm 0.01$ and $0.73\pm 0.02$
 in the regimes II and III, respectively, and they cross at the point $(1,1)$ 
within our numerical accuracy. 
The data of the shortest $M=100$ chain, on the other hand,
deviates from the others in early time region $t/\tau_\theta\lesssim 2\times10^{-1}$, 
where it is in the regime I.
So it can be said  that dynamics in regimes II and III for $M\ge 200$ is interpreted 
as monomer motion through blobs along the self-avoiding tube. 
%\begin{wrapfigure}[20]{i}[0.pt]{8.cm}
%    \scalebox{0.44}{\includegraphics{T20GTM.NT.ps}}
%    \caption{
%The double logarithmic plot of $\phi_{M/2}(t/\tau_\theta)/aR_I$ with $t/\tau_\theta$. 
%The exponents of auxiliary lines are $3/8$ and $3/4$ respectively. 
%}
%    \label{fig:T20GTM.NT}
%\end{wrapfigure}

\begin{figure}
\leavevmode\epsfxsize=80mm
\epsfbox{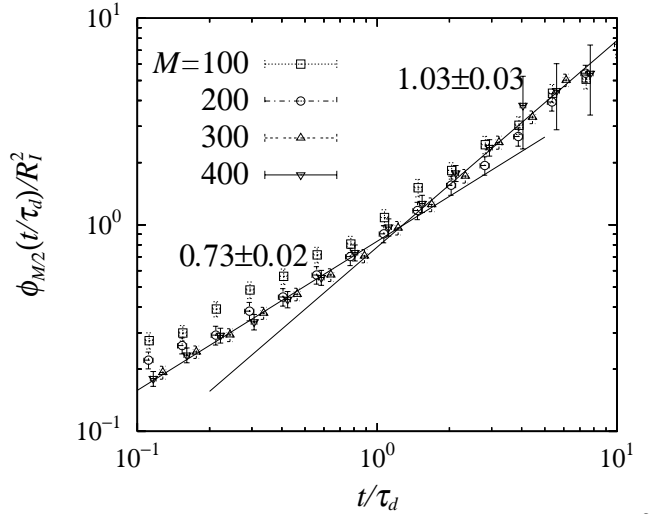}
%\epsfbox{scl3-4.ps}
\caption{
The double logarithmic plot of $\phi_{M/2}(t/\tau_d)/R_I^2$ with $t/\tau_d$. 
}
\label{fig:scl3-4}
\end{figure}

We show double logarithmic plots of $\phi_{M/2}(t/\tau_d)/R_I^2$ against
$t/\tau_d$ in regimes III and IV in Fig.~\ref{fig:scl3-4}. 
Here $\tau_d=\tau_e\left[R_I/a_{\rm b}\right]^{3/\nu_F} \ (\propto 
M^3)$. The data in each regime are scaled to a single curve for $M\ge 200$, and
 they cross near $(1,1)$ as well except for the shortest ($M=100$) chain.
This result supports the crossover 
scenario from the reptation regime of III to the over-all diffusion 
regime of IV. 
%\begin{wrapfigure}[20]{i}[0.pt]{8.cm}
%    \scalebox{0.44}{\includegraphics{T20GTM.ND.ps}}
%    \caption{
%The double logarithmic plot of $\phi_{M/2}(t/\tau_d)/R_I^2$ with $t/\tau_d$. 
%The exponents of auxiliary lines are $3/4$ and $1$ respectively. 
%}
%    \label{fig:T20GTM.ND}
%\end{wrapfigure}

\begin{figure}
\leavevmode\epsfxsize=80mm
\epsfbox{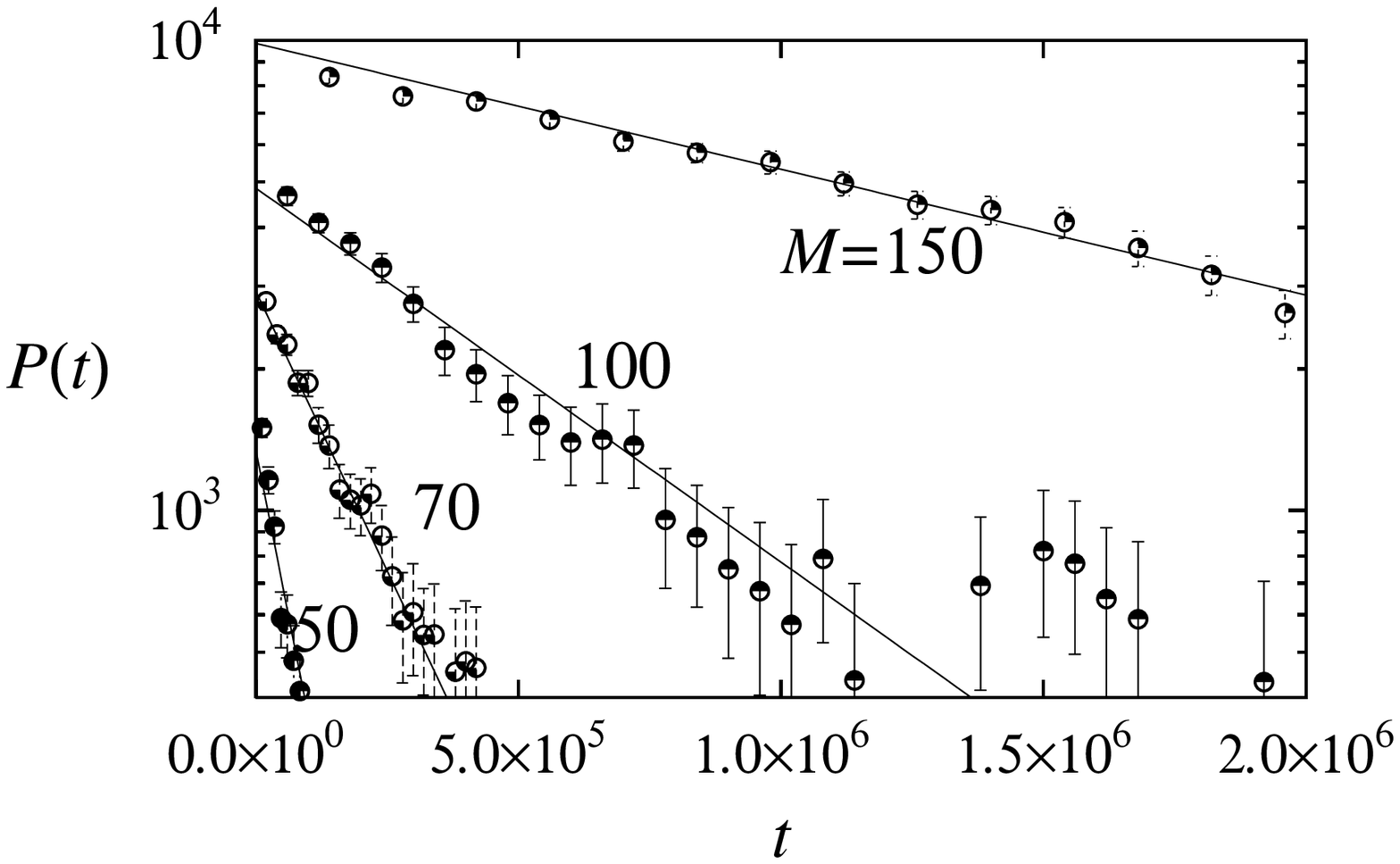}

\leavevmode\epsfxsize=80mm
\epsfbox{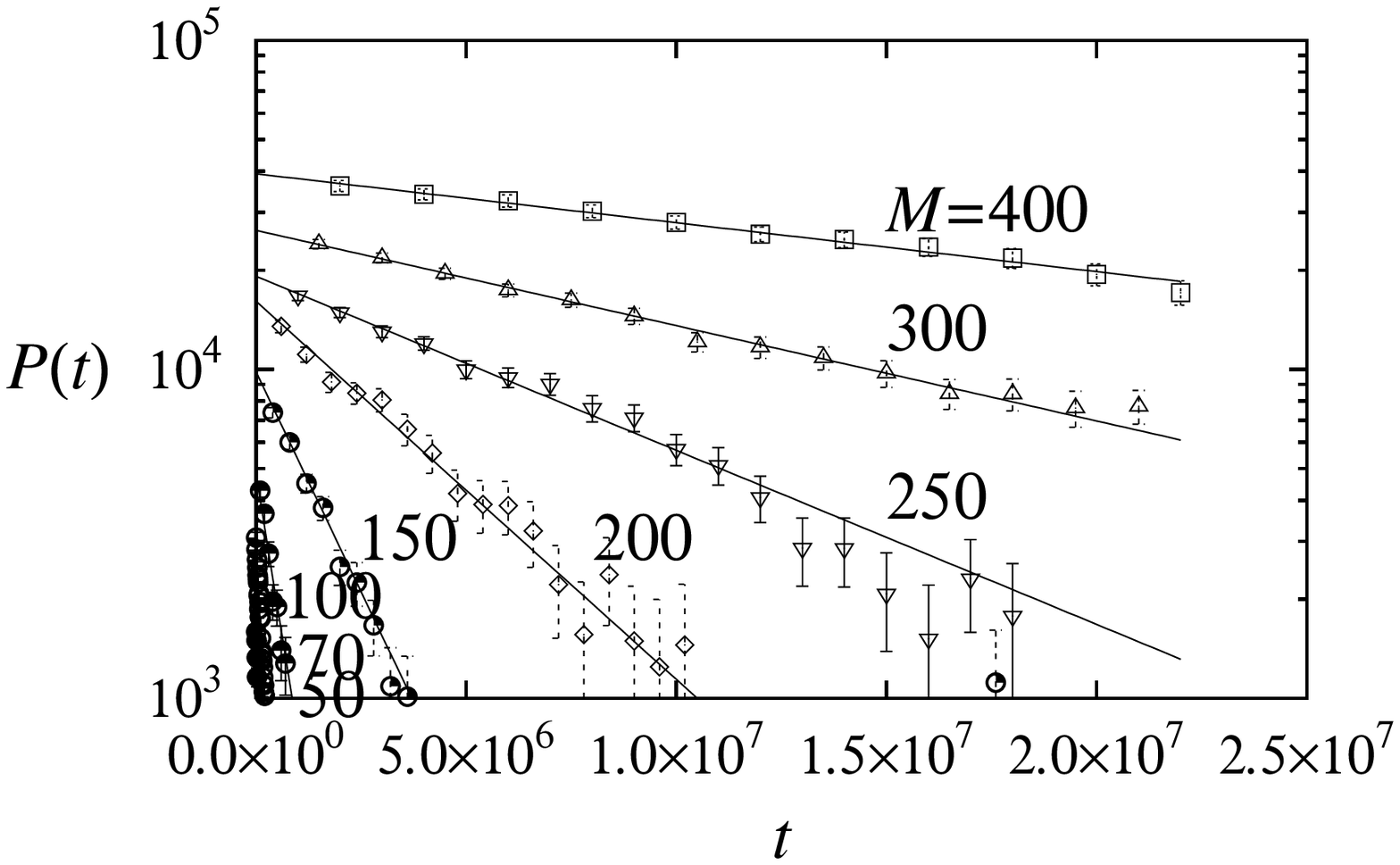}
%\epsfbox{end-end.ps}
\caption{
The autocorrelation function $P(t)$ of the end-to-end vector observed for 
each lengths. 
}
\label{fig:end-end}
\end{figure}

\begin{figure}
\leavevmode\epsfxsize=80mm
\epsfbox{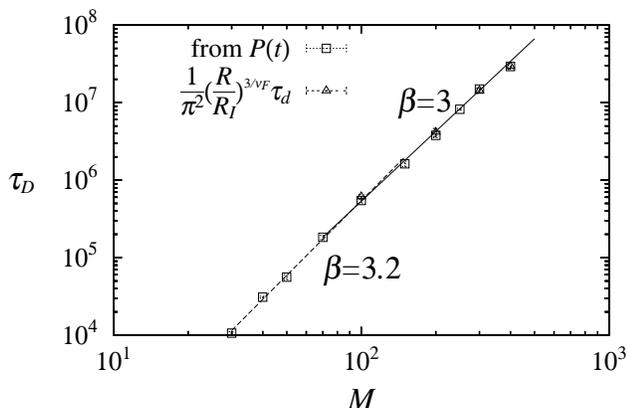}
%\epsfbox{reptat-t.ps}
\caption{
The rotational relaxation times $\tau_D$ obtained from Fig.~\ref{fig:end-end}
and $\frac{1}{\pi^2}\left[\frac{R}{R_I}\right]^{\nu_F/3}\tau_d$ of eq.~(\ref{eqn:tau_dD}).
}
\label{fig:reptat-t}
\end{figure}

Next we discuss behavior of some other quantities than $\phi_{M/2}(t)$ mainly
 in regime IV. 
Figure~\ref{fig:end-end} shows the semi-log plot of $P(t)$ 
for $M=50 \sim 400$. At long time regions, which corresponds to regime IV 
with $\phi_{M/2}(t)\propto t$, $P(t)$ of each $M$ is well fitted to 
an exponential function. 
The plots of the relaxation time extracted from $P(t)$ and the one evaluated by means of
 the r.h.s. of eq.~(\ref{eqn:tau_dD}) against $M$ are shown in Fig.~\ref{fig:reptat-t}. 
The two estimates of $\tau_D$  coincide with each other even up to numerical
 coefficient and clearly display the slope 3
 within error of $3\%$, i.e. $3.0\pm 0.1$, for $M\ge 70$. The result strongly
 supports the argument that
the primitive chain, whose unit length is identical to blob size $a_{\rm b}$,
diffuse along itself. 
%???????----> Although there may be some effect of finite size condition, 
%the range of $M$ is too large to see such correction. 
%So it can be said that the result show the validity of reptation in the system. <-----?????????

%\begin{wrapfigure}[20]{i}[0.pt]{8.cm}
%    \scalebox{0.44}{\includegraphics{T20GTM.RRR-all.ps}}
%    \caption{
%The autocorrelation function $P(t)$ of the end-to-end vector observed for each %lengths. 
%}
%    \label{fig:T20GTM.RRR-all}
%\end{wrapfigure}

\begin{figure}
\leavevmode\epsfxsize=80mm
\epsfbox{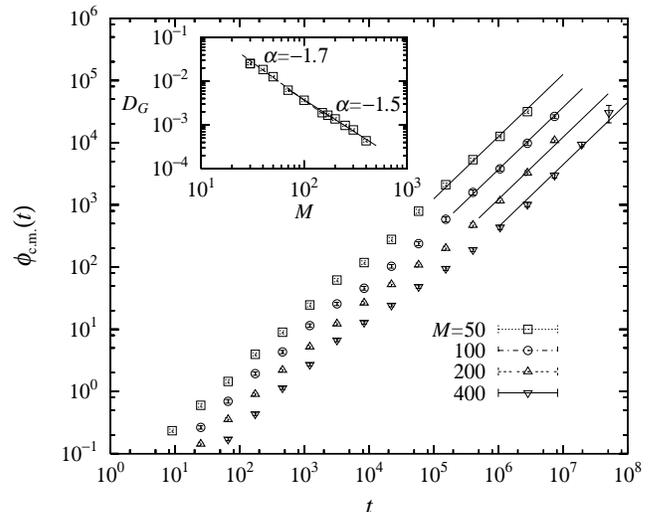}
%\epsfbox{diffus-c.ps}
\caption{
The mean square displacements of the center of mass $\phi_{\rm c.m.}(t)$ and 
their diffusion constants $D_G$, calculated from tangent of final linear regime,
 are plotted against $M$. 
}
\label{fig:diffus-c}
\end{figure}

%\begin{wrapfigure}[20]{i}[0.pt]{8.cm}
%    \scalebox{0.44}{\includegraphics{T20GTM.RRR.ps}}
%    \caption{
%The rotational relaxation times $\tau_d$ obtained from Fig.~\ref{fig:T20GTM.RRR-all}. 
%}
%    \label{fig:T20GTM.RRR}
%\end{wrapfigure}
In such a free diffusion regime monomer motion is identical to that of the center of mass. 
The pre-factor of linear fit of the mean square displacement of the center 
of mass $\phi_{\rm c.m.}(t)$ in the longest time range (which includes 
regime IV) gives us diffusion constant $D_G$~\cite{DoiEdwa}. Typical data 
of $\phi_{\rm c.m.}(t)$ are shown in Fig.~\ref{fig:diffus-c}. We have 
determined $D_G$ by the linear fit of the data in $t \gtrsim \tau_d$, and 
plotted them against $M$ in the inset of Fig.~\ref{fig:diffus-c}.
%, the measured $D_G$ are plotted in terms of $M$. 

\begin{figure}
\leavevmode\epsfxsize=80mm
\epsfbox{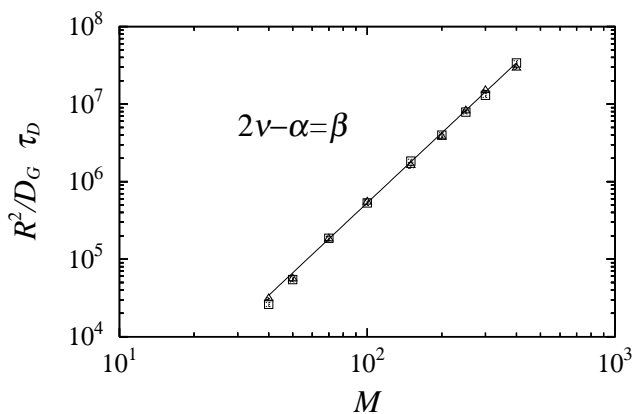}
%\epsfbox{taud-DG.ps}
\caption{
The scaling relation $\beta = 2\nu-\alpha$.
}
\label{fig:taud-DG}
\end{figure}

The exponent $\alpha$ for $D_G\sim M^\alpha$ is estimated as $\alpha\sim -1.5$
 for large $M$. It is different from what Gaussian tube theory 
predicts ($\alpha\sim -2$). It is, instead, readily understood if we plot
$\tau_d$ ( $\propto M^\beta$ ) and $R_I^2/D_G$ ( $\propto M^{2\nu-\alpha}$ ) against
 $M$ simultaneously. As seen in 
Fig.~\ref{fig:taud-DG} the scaling relation $\beta = 2\nu-\alpha$ 
holds as is derived from scaling argument of the reptation 
theory~\cite{deGennes}. 

We have obtained $\alpha \sim -1.5$ using the data with $M=70\sim 400$.
When the data with $M=30\sim 150$ are used, on the other hand, it
becomes $\alpha \sim -1.7$. This is consistent with the results obtained in
our previous work \cite{AT-1} and is quantitatively in agreement with
other results such as reported in \cite{deuma-2}. 
Such a large magnitude of $\alpha$ obtained from smaller $M$ may be
 attributed to tube-length fluctuation which gives rise to a leading
 correction to the scaling described so far. Actually we have confirmed
 $\beta= 3.2\pm 0.1$ from the data with $M=30\sim 150$ as seen in Fig.~\ref{fig:reptat-t}.
Then the scaling relation $\beta=2\nu-\alpha$ still holds for these exponents
 as shown in Fig.~\ref{fig:taud-DG}.

Now we taking into account such correction, 
the deviation of the $M=100$ data from the scaling curve in Fig.~\ref{fig:scl3-4}
is naturally interpreted that $M=100$ is in the crossover range of the parameter space of $M$.

%On the one hand we have obtained $\alpha \sim -1.5$ for $M=70\sim 400$
%but on the other hand it shows $\alpha \sim -1.7$ using
%$D_G$ for $M=30\sim 150$ this is consistent with the result measured in the previous
% work \cite{AT-1} and qualitatively in agreement with other results such as obtained
% in \cite{deuma-2}.
%Such rather steep exponent for the lower $M$ is qualitatively attributed to 
%tube length fluctuation which gives 
%the first term of the correction to rotational relaxation time to the crossover regime
%before reptation regime. 
%The term is a negative dumping function which follows some negative
% and fractional power low dependence to $M$ \cite{DoiEdwa}. 
%As a result the exponent of the relaxation time apparently displays larger value than
%reptation theory expects $\beta > 3$. 
%Actually we could confirm $\beta\sim 3.2$ using rotational relaxation time for $M=30\sim150$
%as seen in Fig. \ref{fig:reptat-t}. 
%Thus using the relation $R_I^2/D_G\propto \tau_d$, $D_G\propto M^{-1.7}$ is derived
%and such relation is confirmed by Fig. \ref{fig:diffus-c}. 

\section{Conclusions}

We investigated on the dynamics of single long self-avoiding polymer in the regularly 
distributed [spacing $a=20$, i.e. concentration $c=1\%$] obstacles in 2D by the e-BFM
 and analyzed the results especially in connection with the tube model. 
The following four regimes are observed in the mean square displacement of center monomer
 $\phi_{M/2}(t)\propto t^{\nu_m}$ as $\nu_m\sim 0.6$, $3/8$, $3/4$ and $1$. 
 These exponents are explained by the ``self-avoiding tube'' model.
That is, the defect dynamics through blobs along such a swelled tube give the exponent
 of $\nu_m$ as $3/8$ and  $3/4$ at the second and the third regimes respectively. 
We have shown the validity of such argument through the scaling analyses on the crossover
 regions and the free diffusion regime. 

%It may be interesting to examine higher dimensions. The exponents easily deduced from
%the arguments in 2 dimensions. We expect $0.54$, $3/10$, $3/5$ and $1$ for $\nu_m$
%in 3 dimensions, and would like to try near future. 
\section*{Acknowledgments}

%We acknowledge useful discussions with . 
The computation in this work has been done using the facilities of 
the Supercomputer Center, Institute for Solid State Physics, 
University of Tokyo, and those of the Computer Center of University 
of Tokyo.

\bibliographystyle{prsty}

%\bibliography{bibs/myrefs}
%\input{figs2.tex}
%\end{multicols}

\end{document}